\documentclass[aip,jcp,amsmath,amssymb,reprint]{revtex4-2}

\usepackage{graphicx}
\usepackage{dcolumn}
\usepackage{bm}
\usepackage[utf8]{inputenc}
\usepackage[T1]{fontenc}
\usepackage{mathptmx}
\usepackage{natbib}
\usepackage{enumitem}   
\usepackage[table,xcdraw]{xcolor}
\usepackage{array}
\usepackage{amsmath, lipsum}

\begin{document}
\preprint{AIP/123-QED}
\title{5- and 6-membered rings: A natural orbital functional study}

\author{Ion Mitxelena}
\email{ion.mitxelena@ehu.eus}
\affiliation{Fisika Aplikatuko departamentua, Vitoria-Gasteiz Ingenieritza Eskola, Euskal Herriko Unibertsitatea (EHU), 01006 Vitoria-Gasteiz, Euskadi, Spain}

\author{Juan Felipe Huan Lew-Yee}
\email{felipe.lew.yee@dipc.org}
\affiliation{Donostia International Physics Center (DIPC), 20018 Donostia, Euskadi, Spain}
\affiliation{Departamento de F\'isica y Qu\'imica Te\'orica, Facultad de Qu\'imica, Universidad Nacional Aut\'onoma de M\'exico, M\'exico City, C.P. 04510, M\'exico}
\affiliation{Departamento de Matem\'aticas, Universidad Nacional Aut\'onoma de M\'exico, M\'exico City, C.P. 04510, M\'exico}

\author{Mario Piris}
\email{mario.piris@ehu.eus}
\affiliation{Donostia International Physics Center (DIPC), 20018 Donostia, Euskadi, Spain}
\affiliation{Polimero eta Material Aurreratuak: Fisika Kimika eta Teknologia, Euskal Herriko Unibertsitatea (EHU), 20018 Donostia, Euskadi, Spain}
\affiliation{Basque Foundation for Science (IKERBASQUE), 48009 Bilbao, Euskadi, Spain}

\date{\today}

\begin{abstract}

The Global Natural Orbital Functional (GNOF) provides a straightforward approach to capture most electron correlation effects without needing perturbative corrections or limited active spaces selection. In this work, we evaluate both the original GNOF and its modified variant GNOFm on a set of twelve 5- and 6-membered molecular rings, systems characterized primarily by dynamic correlation. This reference set is vital as it comprises essential substructures of more complex molecules. We report complete-basis-set limit correlation energies for GNOF, GNOFm, and the benchmark CCSD(T) method. Across the Dunning basis sets, both functionals deliver a balanced and accurate description of the molecular set, with GNOFm showing small but systematic improvements while preserving the overall robustness of the original formulation. These results confirm the reliability of the GNOF family and its ability to capture dynamic correlation effects.

\end{abstract}

\maketitle

\section{Introduction}
While the emergence of deep-learning and similar techniques has led to an improvement of parametrized methods, there is still room for \textit{ab-initio} modern electronic structure methods. The latter are the unique alternative to practice discovery science, as recently shown by J. J. Eriksen \textit{et al.}\cite{eriksen-benzene}, who established the ground-state of Benzene by means of a blind challenge. However, emerging electronic structure methods require benchmarking, not only as a tool for comparison but a necessary test for validation. Benchmark studies provide a quantitative measure of the errors introduced by an approximation in computing different observables, which is essential for assessing the reliability of new approaches. Benzene molecule has been lately exploited as a test system for new methods \cite{Lee-benzene,Loos-cipsi-benzene,greiner-eriksen-benzene}. Damour and co-workers \cite{damour-rings-2021} extended the aforementioned study of Benzene to a 12 molecular set compound by five- and six-membered rings. They investigated the performance and convergence properties of popular single-reference approaches, such as the M{\o}ller-Plesset perturbation series and the coupled-cluster (including iterative approximations) series, in comparison with full configuration interaction (FCI) correlation energy estimates. More importantly, the set included simple aromatic rings form the basis of more complex molecules of biological interest, so an accurate description is desired before going for larger and more complex systems. The motivation of the present study is to employ this molecular set to validate the performance of recent Natural Orbital Functional (NOF) approaches on molecules predominantly dynamic in correlation character.

NOF theory (NOFT),\cite{Piris2007} as the one-particle reduced density matrix (1RDM) functional theory \cite{Gilbert1975, Levy1979, Valone1980, Schilling2018,Schilling2019} in the natural orbital representation,\cite{Lowdin1955, Davidson1972} along with other reduced density matrix methods,\cite{Mazziotti2007, deprince-tutorial} bridges the gap between DFT and wavefunction methods. Unlike the latter, which suffer from steep computational scaling, NOFT achieves a more efficient fifth-order scaling, reducible to fourth-order,\cite{Lew-Yee2021} while accurately describing correlated electronic states. By utilizing the 1RDM and appropriately reconstructing the two-particle reduced density matrix (2RDM) from it, NOFT shows strong potential as a reliable alternative for multireference systems. Today, the complete active space self-consistent field (CASSCF)\cite{Roos1980, Roos2007} approach and its combination with second-order perturbation theory (CASPT2) \cite{Roos1982, Andersson1990, Andersson1992, Pulay2011} remain the most reliable options. However, two major limitations significantly restrict the applicability of CASSCF and CASPT2: the need for active space selection and the high computational cost associated with a large number of strongly correlated orbitals. In contrast, NOF calculations correlate all electrons across all available orbitals within a given basis set, eliminating the complexities of active space selection. This makes NOFT particularly well-suited for problems such as bond-breaking and bond-forming reactions \cite{Piris2024b,Lew-Yee2025c}, where a predefined active space may not be optimal. Additionally, the absence of user-defined input parameters removes arbitrariness and simplifies calculations, making NOFs more accessible to non-experts and appropriate to carry out studies without prior knowledge of the system, \textit{e.g.} blind challenges.

Over the past two decades, NOFT has advanced significantly from both theoretical and computational perspectives. On the theoretical side, Piris and co-workers have developed a family of functionals known as PNOFs,\cite{Piris2013b, Piris2014c, Piris2017, Mitxelena2018a, Piris2021a} which continue to demonstrate their competitiveness with standard electronic structure methods. Their capabilities extend to various domains, including the description of excited states\cite{Lew-Yee2024} and molecular dynamics,\cite{RiveroSantamaria2024} as well as significant advancements in mitigating delocalization errors,\cite{Lew-Yee2023b} a persistent challenge in DFT. Additionally, PNOFs have contributed to understanding the ground-state spin state of iron(II) porphyrin,\cite{Lew-Yee2023a} a long-standing problem in electronic structure theory. More recently, NOFs have been employed for energy measurements on quantum computers, significantly improving efficiency within the variational quantum eigensolver (VQE) framework, giving rise to NOF-VQE.\cite{Lew-Yee2025a}

On the computational side, while NOFT calculations were initially constrained by high computational costs, recent advances have significantly improved their efficiency.\cite{franco_softmax, Lew-Yee2025b} A key development in this direction has been the incorporation of modern numerical techniques inspired by deep learning,\cite{Lew-Yee2025b} particularly momentum-based optimization methods such as the ADAM optimizer, which have accelerated the convergence of natural orbital calculations. These improvements have enabled NOFT to handle strongly correlated systems with up to 1000 electrons, the largest NOF calculations to date, making NOFT a viable tool for large-scale applications.

Despite these advances, NOFT remains underutilized, primarily due to two factors. First, NOF methods are not yet implemented in widely used electronic structure software packages. Although the open-source DoNOF program\cite{Piris2021,lew-yee2026} for NOF calculations represents a significant step forward, broader integration is still needed. Second, accessible and systematic assessments of NOFs’ performance are scarce, making it difficult for researchers to gauge its reliability. In this vein, while the aforementioned GNOF approximation has been tested on strongly correlated models,\cite{Mitxelena2022,Mitxelena2024} its accuracy on systems dominated by dynamic correlation is undetermined yet, so a step forward in this direction is intended in the present work.

This article is organized as follows. The basics of NOFT are described in next section \ref{sec:theory}, as well as the electron-pairing-based GNOF approximation and its modification GNOFm employed later on. In section \ref{sec:theory2}, the system set is introduced together with the methods that are used to compare with. Then, GNOF and GNOFm results are presented in section \ref{sec:results}, together with reference CCSD(T) calculations. The article ends with a few remarks in section \ref{sec:closing}.

\section{Electron-pairing-based NOFs}\label{sec:theory}

In this section, we outline the key concepts of NOFT to clarify its differences from commonly used approaches for studying strongly correlated systems. A more detailed description of NOFT and the approximations that define different NOFs can be found in Ref.~[\citenum{Piris2024}]. Additionally, Ref.~[\citenum{Piris2024a}] presents a perspective on NOFT, discussing its fundamental concepts, strengths and weaknesses, current status, and potential future developments.

The energy of any NOF is typically expressed in terms of the set of NOs $\{\phi_i\}$ and their ONs $\{n_i\}$ as
\begin{equation}
    E[\mathrm{N},\{n_i,\phi_i\}]= \sum _i n_i H_{ii} + \sum  _{ijkl} D[n_i,n_j,n_k,n_l] \langle ij|kl \rangle 
    \label{Enoft}   
\end{equation}
where the one- and two-electron integrals in the NO basis are given by
\begin{equation}
H_{ii} =\int d{\bf r} \phi _i ^*({\bf r})\left(-\frac{\nabla ^2 _r} {2}+v({\bf r}) \right) \phi _i ({\bf r})
\label{hpp}   
\end{equation}
\begin{equation}
    \langle ij|kl \rangle= \int \int d{\bf r}_1 d {\bf r}_2 \frac{\phi ^* _i ({\bf r}_1)\phi ^* _j ({\bf r}_2)\phi _k ({\bf r}_1)\phi _l({\bf r}_2)}{|{\bf r}_2 -{\bf r}_1|}
    \label{eri} 
\end{equation}
In Eq. (\ref{hpp}), $v({\bf r})$ represents the nuclear potential determined by molecular geometry within the Born-Oppenheimer approximation, assuming no additional external fields. Unlike DFT, NOFT does not require a reconstruction for the one-electron part. However, the explicit form of the electron-electron interaction energy functional remains unknown, and different functional forms of $D[n_i,n_j,n_k,n_l]$ lead to distinct NOFs. 

The approximate functional (\ref{Enoft}) explicitly depends on the 2RDM,\cite{Donnelly1979} requiring not only the N-representability of the 1RDM\cite{Coleman1963} but also that of the functional itself.\cite{Piris2018d} Specifically, the reconstructed $D[n_i,n_j,n_k,n_l]$ must satisfy the same N-representability conditions as an unreconstructed 2RDM\cite{Mazziotti2012} to ensure the existence of a compatible N-electron system. Given their implicit dependence on the 2RDM, approximate functionals are best classified as NOFs rather than pure 1RDM functionals, as they are only defined in the NO representation.

In this article, we focus on electron-pairing-based functionals, which have proven particularly effective for describing strongly correlated systems and offer significant advantages from both theoretical and practical perspectives.\cite{Piris2018e} Accordingly, we consider $\mathrm{N_{I}}$ unpaired electrons that determine the system's total spin $S$, while the remaining $\mathrm{N_{II}} = \mathrm{N-N_{I}}$ electrons form pairs with opposite spins, resulting in a net spin of zero for the $\mathrm{N_{II}}$ electrons. 

We focus on the highest-multiplicity mixed state, where $2S+1=\mathrm{N_{I}}+1$ and the expectation value of $\hat{S}_{z}$ is zero. Consequently, the spin-restricted formalism can be applied, ensuring that all spatial orbitals $\{\varphi_{p}\}$ are doubly occupied within the ensemble and that $\alpha$ and $\beta$ spin particles have equal occupancies.\citep{Piris2019}

Following the partitioning of electrons into $\mathrm{N_{I}}$ and $\mathrm{N_{II}}$, the orbital space $\Omega$ is divided into two subspaces: $\Omega = \Omega_{\mathrm{I}} \oplus \Omega_{\mathrm{II}}$. The subspace $\Omega_{\mathrm{II}}$ is composed of $\mathrm{N_{II}}/2$  mutually disjoint subspaces $\Omega{}_{g}$, each containing a reference orbital $\left|g\right\rangle$ for $g\leq\mathrm{N_{II}}/2$, along with $\mathrm{N}_{g}$ associated orbitals $\left|p\right\rangle $ for $p>\mathrm{N_{II}}/2$, formally expressed as
\begin{equation} 
\Omega_{g}=\left\{ \left|g\right\rangle ,\left|p_{1}\right\rangle ,\left|p_{2}\right\rangle ,...,\left|p_{\mathrm{N}_{g}}\right\rangle \right\} .\label{OmegaG}
\end{equation}
Considering spin, the total occupancy of a given subspace $\Omega_{g}$ is 2, as expressed by the following pairing condition:
\begin{equation}
\sum_{p\in\Omega_{g}}n_{p}=n_{g}+\sum_{i=1}^{\mathrm{N}_{g}}n_{p_{i}}=1,\quad g=1,2,...,\frac{\mathrm{N_{II}}}{2}.\label{sum1}
\end{equation}
Similarly, $\Omega_{\mathrm{I}}$ consists of $\mathrm{N_{I}}$ mutually disjoint subspaces $\Omega_{g}$. Unlike $\Omega_{\mathrm{II}}$, each subspace $\Omega{}_{g} \in \Omega_{\mathrm{I}}$ contains only one orbital $g$ with an ON of $n_{g}=1/2$. Notably, each orbital holds a single electron, though its specific spin state, whether $\alpha$ or $\beta$, remains undetermined. From Eq. (\ref{sum1}), it follows that the trace of the 1RDM equals the total number of electrons:
\begin{equation}
2\sum_{p\in\Omega}n_{p}=2\sum_{p\in\Omega_{\mathrm{II}}}n_{p}+2\sum_{p\in\Omega_{\mathrm{I}}}n_{p}=\mathrm{N_{II}}+\mathrm{N_{I}}=\mathrm{\mathrm{N}}.\label{norm}
\end{equation}
The simplest electron-pair-based functional is PNOF5, which describes independent electron pairs,\cite{Piris2011, Piris2013a} and its energy expression is given by:
\begin{equation}
E\left[\mathrm{N},\left\{ n_{p},\varphi_{p}\right\} \right] = E^\text{intra}+E_\mathrm{HF}^\text{inter} \label{pnof5}
\end{equation}

The intra-pair component is formed by summing the energies $E_{g}$ of electron pairs with opposite spins and the single-electron energies of unpaired electrons, specifically,
\begin{equation}
E^\mathrm{intra}=\sum\limits _{g=1}^{\mathrm{N_{II}}/2}E_{g}+{\displaystyle \sum_{g=\mathrm{N_{II}}/2+1}^{\mathrm{N}_{\Omega}}} H_{gg}
\label{Eintra}
\end{equation}
\begin{equation}
E_{g} = 2 \sum\limits _{p\in\Omega_{g}}n_{p}H_{pp} + \sum\limits _{q,p\in\Omega_{g}} \Pi(n_q,n_p) L_{pq}
\end{equation}
where $L_{pq}=\left\langle pp|qq\right\rangle$ are the exchange-time-inversion integrals.\cite{Piris1999} In Eq. (\ref{Eintra}), $\mathrm{\mathrm{N}_{\Omega}}=\mathrm{N_{II}}/2+\mathrm{N_{I}}$ denotes the total number of suspaces in $\Omega$. The matrix elements $\Pi(n_q,n_p) = c(n_q)c(n_p)$, where $c(n_p)$ is defined by the square root of the ONs according to the following rule:
\begin{equation}
    c(n_p) = \left.
  \begin{cases}
    \phantom{+}\sqrt{n_p}, & p \leq \mathrm{N_{II}}/2\\
    -\sqrt{n_p}, & p > \mathrm{N_{II}}/2 \\
  \end{cases}
  \right. \>\>
  \label{eq:PNOF5-roots}
\end{equation}
that is, the phase factor of $c(n_p)$ is chosen to be $+1$ for the strongly occupied orbital of a given subspace $\Omega_g$, and $-1$ otherwise. The inter-subspace Hartree-Fock (HF) term is 
\begin{equation}
E_\text{HF}^\text{inter}=\sum\limits _{p,q}^{\mathrm{N}_B}\,'\,n_{q}n_{p}\left(2J_{pq}-K_{pq}\right)\label{ehf}
\end{equation}
where $J_{pq}=\left\langle pq|pq\right\rangle$ and $K_{pq}=\left\langle pq|qp\right\rangle $ are the Coulomb and exchange integrals, respectively. $\mathrm{N}_{B}$ denotes the number of basic functions considered. The prime in the summation indicates that only the inter-subspace terms are taken into account.

To enhance the inter-pair electron correlation, inter-subspace static and dynamic components must be added which lead to GNOF.\cite{Piris2021a} Its corresponding energy expression is given by:
\begin{equation}
E\left[\mathrm{N},\left\{ n_{p},\varphi_{p}\right\} \right] = E^\mathrm{intra} + E_\mathrm{HF}^\mathrm{inter} + E_\mathrm{sta}^\mathrm{inter} + E_\mathrm{dyn}^\mathrm{inter} \label{gnof}\\
\end{equation}
where
\begin{multline}
    E_\mathrm{sta}^\mathrm{inter} = -\bigg( \sum_{p=1}^{N_\Omega}\sum_{q=N_\Omega+1}^{N_\mathrm{B}} +  \sum_{p=N_\Omega+1}^{N_\mathrm{B}}\sum_{q=1}^{N_\Omega} + \sum_{p,q=N_\Omega+1}^{N_\mathrm{B}} \bigg)' \Phi_q \Phi_p L_{pq} \\
    -\frac{1}{2}\bigg( \sum_{p=1}^{N_\text{II}/2}\sum_{q=N_\text{II}/2+1}^{N_\Omega} + \sum_{p=N_\text{II}/2+1}^{N_\Omega}\sum_{q=1}^
    {N_\text{II}/2} \bigg)' \Phi_q \Phi_p L_{pq} \\
    - \sideset{}{'}\sum_{p,q=N_\text{II}/2+1}^{N_\Omega} \Phi_q \Phi_p K_{pq}
    \label{esta7}
\end{multline}
\begin{equation}
E_\text{dyn}^\text{inter}= \sum\limits _{p,q=1}^{\mathrm{N}_B}{''\>}
\left[\Pi(n_{q}^{d},n_{p}^{d})+n_{q}^{d}n_{p}^{d}\right] L_{pq}
\label{edyn}
\end{equation}
Here, $\Phi_{p}=\sqrt{n_{p}h_{p}}$ with $h_{p}=1-n_{p}$ being the hole. The second prime in Eq. (\ref{edyn}) additionally excludes interactions between orbitals below the level $\mathrm{N_{II}}/2$. The dynamic contribution to the ON $n_p$ is defined as
\begin{equation}
n_p^d=n_p \cdot e^{-\left(\dfrac{h_g}{h_c}\right)^{2}},\,\,p\in\Omega_{g}\,,\,\, g=1,2,...,\frac{\mathrm{N_{II}}}{2}.
\label{dyn-on}
\end{equation}
with $h_c = 0.02\sqrt{2}$. The maximum value of $n_p^d$ is approximately 0.012, aligning with Pulay’s criterion, which states that an occupancy deviation of $\thickapprox 0.01$ from 1 or 0 is necessary for a NO to contribute to dynamic correlation.

Recently, a modified version of GNOF, denoted GNOFm, reintroduces the interactions between strongly occupied orbitals in the antiparallel spin blocks, as originally proposed in PNOF7.\cite{Piris2017, Mitxelena2018a} This refinement has shown improved accuracy for describing the singlet triplet energy gaps along the linear n-acene series.\cite{Lew-Yee2025b} Within this framework, and assuming real orbitals so that $L_{pq}=K_{pq}$, the inter-subspace static component takes the following compact form:
\begin{equation}
    E_\text{sta}^\text{inter} = - \sum\limits _{p,q}^{\mathrm{N}_B}\,' \Phi_q \Phi_p K_{pq}
    \label{gnofm}
\end{equation}
The solution is established by optimizing the energy with respect to the ONs and NOs, separately. Therefore, orbitals vary along the optimization process until the most favorable orbital interactions are found. All calculations have been carried out using the DoNOF code \cite{Piris2021,lew-yee2026} and the recently implemented orbital optimization algorithm.\cite{Lew-Yee2025b}

\section{Motivation and Methodology}\label{sec:theory2}
Comparisons between different NOFs are rare in the literature. Notable exceptions include studies on the behavior of various functionals, also beyond the electron-pairing approach, in the Hubbard Hamiltonian model \cite{Mitxelena2017a, Mitxelena2018} and a rigorous assessment of 2RDM approximations that give rise to NOFs, evaluating their capacity to satisfy key properties of the exact functional.\cite{RodriguezMayorga2017} Both comparative studies concluded that the functional N-representability is crucial for obtaining consistent results across different electronic correlation regimes. Consequently, we restrict our analysis to the electron-pairing-based NOFs presented in the previous section that enforce (2,2)-positivity conditions on the 2RDM. \cite{Mazziotti2012}

From a practical perspective, electron-pairing-based NOFs are particularly suited for describing strong correlation effects. In particular, the PNOF7 approximation was proven to be an efficient method for studying the Hubbard model and Hydrogen clusters described by a minimal basis set in one- and two-dimensions.\cite{Mitxelena2020a,Mitxelena2020b} Unfortunately, as recently shown by Lew-Yee and Piris,\cite{Lew-Yee2025b} PNOF7 could fail in molecular systems where dynamic correlation effects are non-negligible, and therefore the GNOF approximation is preferable for such systems. As briefly described in the previous section, GNOF aims to describe all electron correlation effects in a balanced manner, and numerous publications have demonstrated its ability to compete with standard electronic structure methods in different scenarios.\cite{Lew-Yee2023b,Lew-Yee2023a,Lew-Yee2025b,Piris2021a} Previous NOF approaches tried to retrieve dynamic correlation effects by terms of perturbation theory,\cite{Piris2013c,Piris2017,Piris2018b} but including them into the functional itself gives access to correlated NOs and ONs. Nevertheless, while GNOF has been tested on model systems for strong correlation in one-, two- and three-dimensions,\cite{Mitxelena2022,Mitxelena2024} benchmarking its performance in systems dominated by dynamic electron correlation remains undone. In view of the results reported in Ref.~[\citenum{Lew-Yee2025b}], GNOF could be improved in complex correlation situations by a recent modification, so GNOFm is also included in the present work. This comparison, indeed, may help to clarify the delicate balance between dynamic and non-dynamic electron correlation terms in electron-pairing-based NOFs.

\begin{figure}[thb]
\begin{centering}
{\includegraphics[scale=0.18]{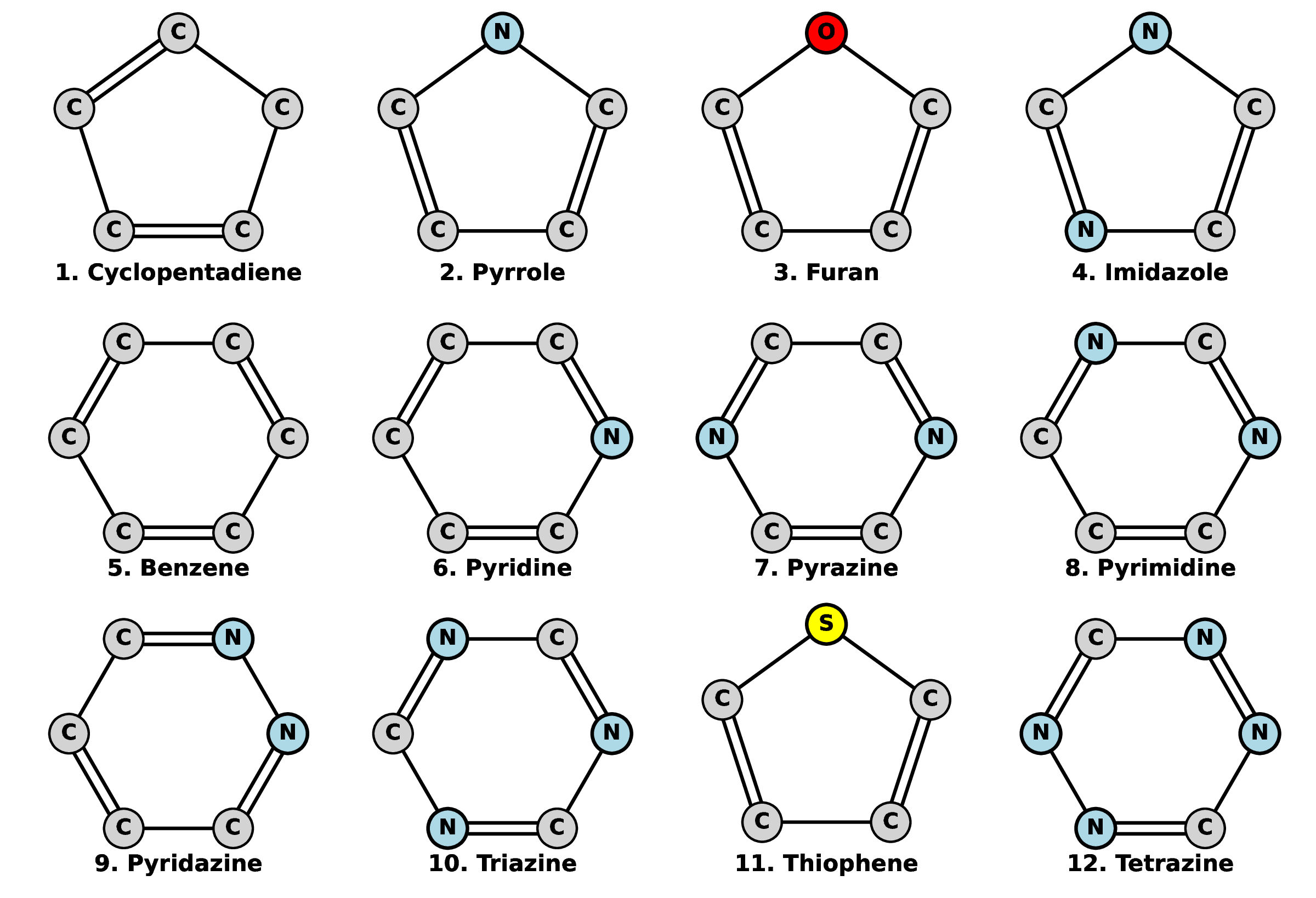}}
\caption{\label{fig:systems} 5- and 6-membered molecular rings studied along this work, as well as the corresponding numbering employed later on.}
\end{centering}
\end{figure}

In Ref.~[\citenum{damour-rings-2021}], Damour \textit{et al.} provided accurate FCI correlation energy estimates for twelve cases of five- and six-membered ring molecules, namely: Cyclopentadiene, Furan, Imidazole, Pyrrole, Thiophene, Benzene, Pyrazine, Pyridazine, Pyridine, Pyrimidine, s-Tetrazine, and s-Triazine.
Hence, the set involves systems with atoms of the first to third lines of the periodic table. An schematic representation of the latter is shown in Fig. \ref{fig:systems}. In particular, Damour and co-workers reported optimized-orbital selected configuration interaction calculations for a correlation-consistent double-$\zeta$ Dunning basis set (cc-pVDZ),\cite{Dunning} as a reference for further studying the convergence of the M\o{}ller-Plesset perturbation theory series and the iterative approximate coupled-cluster series. Even in the context of the cc-pVDZ basis set, computing FCI result of these molecules is too computational demanding. Indeed, today carrying out coupled-cluster with singles, doubles, triples, and quadruples (CCSDTQ) calculations for molecules larger than benzene is prohibitively expensive or at least not practical.\cite{eriksen-benzene} This situation puts NOF approaches in an interesting position to run calculations employing larger basis sets from cc-pVDZ to cc-pV5Z. Nevertheless, as discussed below, all-electron calculations have been employed consistently throughout this work, with no frozen electrons. Therefore, basis sets including weighted core-valence functions are preferred. To this end, we have included Dunning's weighted core-valence basis sets cc-pwCVXZ \cite{dunning-ccpwcvnz}, whereas results obtained by using the cc-pVXZ basis set family are shown in the Supplementary Information. The latter are carried out for X=2-5 cardinal numbers, however, the X=5 level is omitted from the cc-pwCVXZ calculations due to the marked increase in the number of basis functions \cite{helgaker-extrapolation} and the computational cost. In the following, we use these molecular sets to study GNOF and GNOFm correlation energies and their convergence with the size of the basis set. We provide complete-basis-set (CBS) estimates for these approximations, as well as for the ground-state gold standard coupled-cluster singles, doubles, and perturbative triples CCSD(T). Following the work by Damour and co-workers, geometries of the molecular systems, obtained at the CC3/aug-cc-pVTZ level of theory, were extracted from Ref. [\citenum{loos-2020}]. 

The DoNOF code \cite{Piris2021,lew-yee2026} was employed for GNOF and GNOFm calculations, whereas CCSD(T) calculations were carried out with the PSI4 software package.\cite{psi4} In contrast to Damour \textit{et al.}, no frozen core orbitals were considered in the present study. All electrons are correlated through all orbitals given in the basis set within the NOFT framework. The latter is, indeed, a strength of NOFs and their actual advantage with respect to typically used methods for multireference correlation, which require to define an active space where electrons are correlated. Finally, the resolution of identity approximation was used for integral evaluation in NOF calculations. \cite{Lew-Yee2021} The latter was also employed in CCSD(T) calculations through the density-fitting option available in the PSI4 code \cite{psi4}. As demonstrated\cite{deprince-2013} by DePrince III and Sherill, this would affect CCSD(T) energies, at most, in the order of a few $mE_h$, so in any case it alters neither the reported results nor the obtained conclusions.

\section{Results}\label{sec:results}
In this section, we analyze GNOF and GNOFm correlation energies for the aforementioned set of molecules, and compare them with CCSD(T) calculations. Note that correlation energies refer to the difference between energies given by a correlated method $E_\mathrm{M}$ and the Hartree-Fock energies $E_\mathrm{HF}$, i.e. $E_\mathrm{corr} = E_\mathrm{M}-E_\mathrm{HF}$. A summary of raw correlation energies is publicly available in Ref.~[\citenum{data-OSF}].

\begin{figure}[thb]
\begin{centering}
{\includegraphics[scale=0.27]{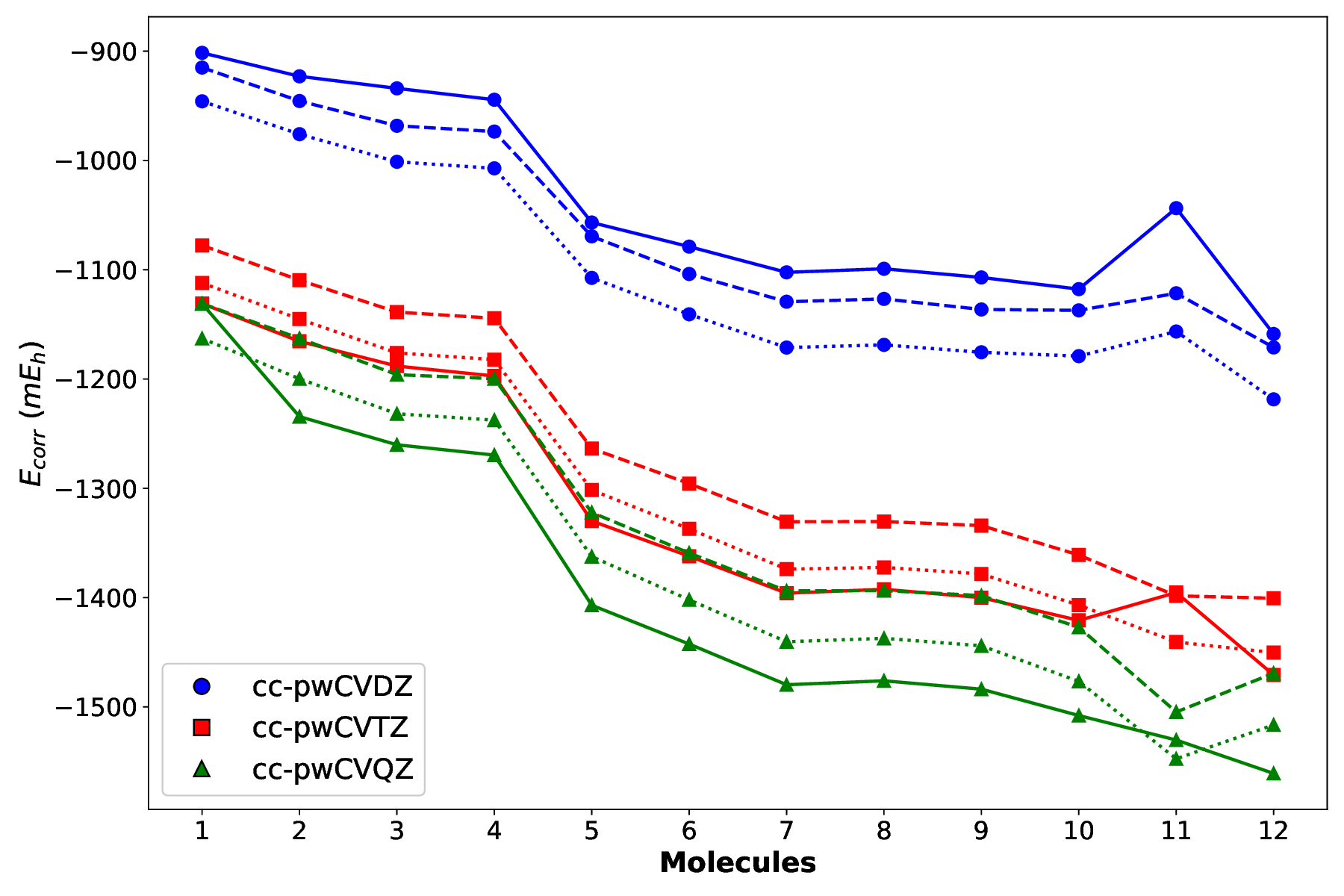}}
\caption{\label{fig:Ecorr_ccd-cc4} Correlation energies ($E-E_\mathrm{HF}$) in mE$_{h}$ for the selected molecules obtained with GNOF (dashed), GNOFm (dotted), and CCSD(T) (solid) using cc-pwCVXZ basis sets. The cardinal numbers X = 2, 3, and 4 correspond to blue, red, and green, respectively. Molecules ordered according to the numbering given in Fig. \ref{fig:systems}.}
\end{centering}
\end{figure}

Correlation energies for GNOF, GNOFm, and CCSD(T) are shown in Fig. \ref{fig:Ecorr_ccd-cc4} for increasing size weighted core-valence Dunning basis sets (cc-pwCVXZ, X=2-4). Here, molecules are ordered from smaller to larger correlation energies, according to the numbering presented in Fig.~\ref{fig:systems}. In the case of thiophene, including $d$ functions in the basis sets is required to obtain a correct description due to the presence of a sulfur atom \cite{BELL-basis-N+D}. In terms of correlation energy, GNOF retrieves 1095 mE$_{h}$ and 1122 mE$_{h}$ for thiophene using the cc-pwCVDZ and aug-cc-pwCVDZ basis sets, respectively. However, this difference reduces to a few mE$_{h}$ for larger basis sets. Concretely, GNOF thiophene correlation energies read as 1397 mE$_{h}$ and 1398 mE$_{h}$ for cc-pwCVTZ and aug-cc-pwCVTZ, respectively, and 1500 mE$_{h}$ and 1505 mE$_{h}$ for cc-pwCVQZ and aug-cc-pwCVQZ, respectively.

A first look to Fig. \ref{fig:Ecorr_ccd-cc4} reveals that NOF and CCSD(T) curves are roughly parallel, represented by dashed and solid lines, respectively, so the molecular description agrees for both methods independently of the size of the basis set, as well as of the different studied molecular rings. In detail, results corresponding to the smallest basis set employed, cc-pwCVDZ, show that GNOF and GNOFm retrieve more correlation energy than CCSD(T), represented by the solid blue line. Nevertheless, this result is reversed when the basis set size is increased. In the case of cc-pwCVTZ and cc-pwCVQZ calculations, represented by red and green lines, respectively, NOF correlation energies remain above CCSD(T) results. GNOFm shows a better agreement with CCSD(T) than its predecessor GNOF in both cases. This agreement is particularly accurate when using the cc-pwCVTZ basis set. However, CCSD(T) correlation energies get larger than GNOFm ones when the basis set increases from cc-pwCVTZ to cc-pwCVQZ, and thereby GNOFm differences with respect to CCSD(T) grow with the basis set. For the sake of completeness, calculations carried out with the Dunning's cc-pVXZ basis sets, $X=2,3,4,5$ being the cardinal number of the basis set, have been included in the Supplementary Information.

\begin{table}[htbp]
\centering
\setlength{\arrayrulewidth}{0.5mm} 
\setlength{\tabcolsep}{6pt} 
\renewcommand{\arraystretch}{1.7} 
\begin{tabular}{|c|c|c|c|c|}
\hline
No. & \textbf{Systems} & \textbf{GNOF} & \textbf{GNOFm} & \textbf{CCSD(T)} \\ \hline
\hline
1 & Cyclopentadiene & -1156.1 & -1189.5 & -1187.6 \\ \hline
2 & Pyrrole & -1188.2 & -1226.0 & -1274.7 \\ \hline
3 & Furan & -1221.9 & -1259.2 & -1302.5 \\ \hline
4 & Imidazole & -1226.2 & -1265.1 & -1311.5 \\ \hline
5 & Benzene & -1353.4 & -1393.1 & -1452.5 \\ \hline
6 & Pyridine & -1388.5 & -1432.1 & -1489.6 \\ \hline
7 & Pyrazine & -1425.6 & -1471.6 & -1528.4 \\ \hline
8 & Pyrimidine & -1425.9 & -1469.1 & -1524.7 \\ \hline
9 & Pyridazine & -1428.8 &-1475.5  & -1532.3 \\ \hline
10 & Triazine & -1463.6 & -1513.1 & -1558.0 \\ \hline
11 & Thiophene & -1542.9 & -1587.3 & -1579.2 \\ \hline
12 & Tetrazine & -1506.4 & -1554.9 & -1612.3 \\ \hline
\end{tabular}
\caption{\label{table1} Complete basis set (CBS) extrapolated correlation energies ($E-E_\mathrm{HF}$) in mE$_{h}$ for the 12 molecular systems, computed using GNOF, GNOFm, and CCSD(T). Helgaker's extrapolation scheme, $E_{\infty} + bX^{-3}$, was employed with $X = 2, 3, 4$ as the cardinal number of the basis set.}
\end{table}

As shown in Fig.~\ref{fig:Ecorr_ccd-cc4}, energy differences for the same method and molecule decrease as the size of the basis sets is augmented. This suggests that we are approaching the CBS limit for the reported correlation energies. Therefore, we computed the CBS limit correlation energies for the twelve molecules set using GNOF and GNOFm, as well as CCSD(T). Previous studies \cite{MATXAIN2010} demonstrated similar results for different extrapolation schemes within the NOFT framework, concretely, exponential function like: $E(X)=E_{\infty} + A exp(-\gamma X)$ and power function like: $E(X)=E'_{\infty} + A' X^{-\gamma'}$, being $X$ the cardinal number of the basis set. Nevertheless, in this work we employ the Helgaker's extrapolation scheme \cite{helgaker-extrapolation} of the form $E_{corr} = E^{''}_{\infty} + bX^{-3}$, which is considered a standard method today and is particularly adequate for Dunning correlation-consistent basis sets. The corresponding GNOF and GNOFm CBS estimated values are given in Table \ref{table1}, which also presents extrapolations CCSD(T), performed using the same procedure. An inspection of CBS limit molecular correlation energies reveals an agreement within 100 mE$_{h}$ for GNOF in most cases, values that are even improved to around 50 mE$_{h}$ when GNOFm is utilized.

\begin{figure}[thb]
\begin{centering}
{\includegraphics[scale=0.27]{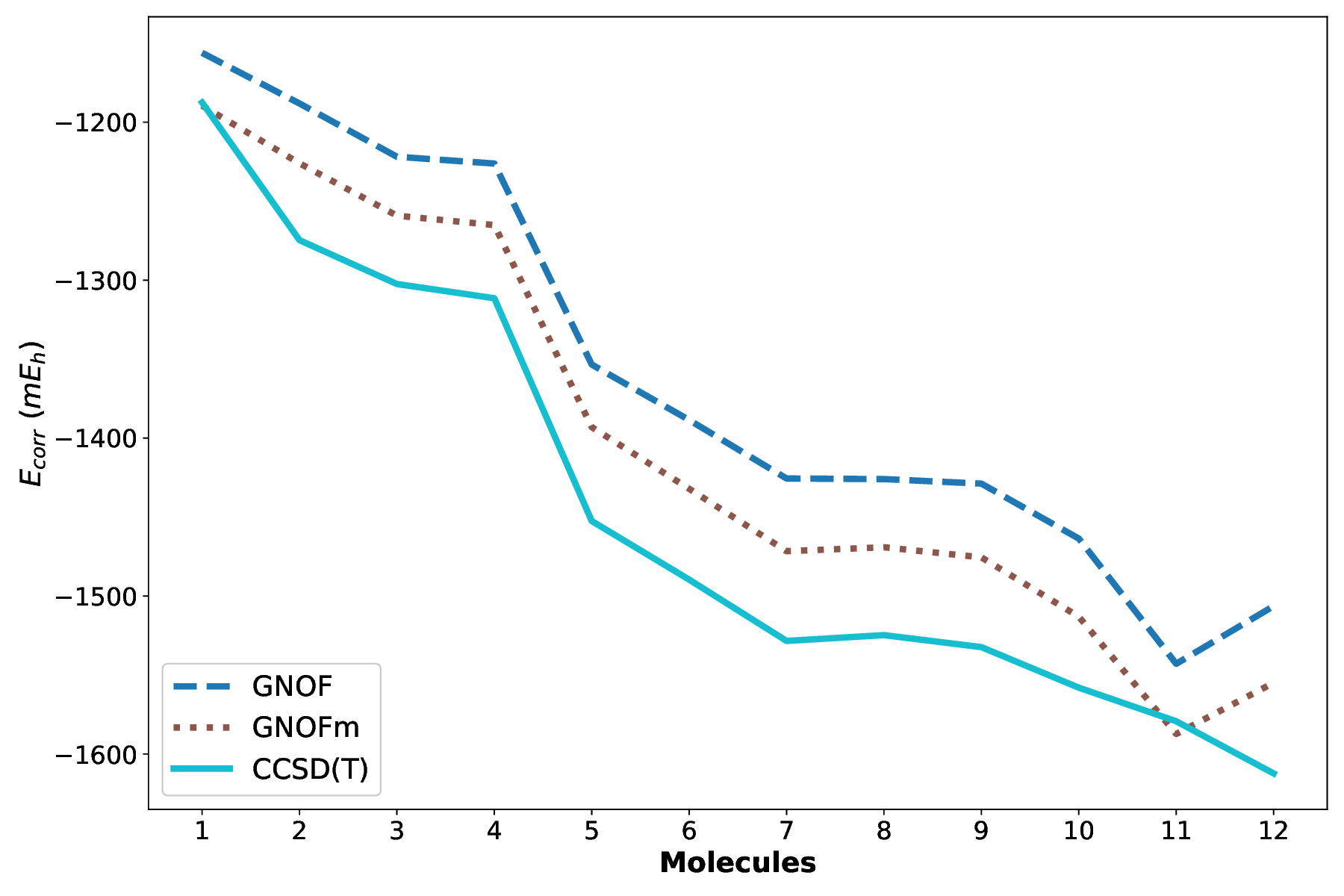}}
\caption{\label{fig:Ecorr_cbslimit} Complete basis set (CBS) extrapolated correlation energies ($E-E_\mathrm{HF}$) in mE$_{h}$ for the 12 molecular systems, computed using GNOF, GNOFm, and CCSD(T). Helgaker's extrapolation scheme, $E_{\infty} + bX^{-3}$, was employed with $X = 2, 3, 4$ as the cardinal number of the basis set.}
\end{centering}
\end{figure}

The results shown in Table \ref{table1} are summarized in Fig. \ref{fig:Ecorr_cbslimit}. Here, we plot CBS extrapolated correlation energies for the 12 molecular systems. GNOFm energies systematically get closer to CCSD(T) results when the static term between electron pairs is modified according to Eq. (\ref{gnofm}). In other words, Fig. \ref{fig:Ecorr_cbslimit} reveals that GNOFm CBS correlation energies reduce differences between GNOF and CCSD(T) to the half.

Both GNOF and its recent modification GNOFm provide qualitatively good descriptions of the five- and six-membered rings, as their traces lie close to and largely parallel with the CCSD(T) reference. Notably, Fig. \ref{fig:Ecorr_cbslimit} also shows a quantitative improvement of GNOFm over GNOF, consistent with Lew-Yee and Piris.\cite{Lew-Yee2025b} More importantly, the figure demonstrates that NOFs recover dynamic-correlation effects across the entire family of correlation-consistent Dunning basis sets and for all systems considered, proving the robustness of the Global NOF approach. Unlike earlier approximations, particularly among electron-pairing based NOFs described in section \ref{sec:theory}, these functionals incorporate dynamic correlation within the energy expression itself and therefore do not require perturbative corrections.

\begin{table}[htbp]
\centering
\setlength{\arrayrulewidth}{0.5mm} 
\setlength{\tabcolsep}{6pt} 
\renewcommand{\arraystretch}{1.7} 
\begin{tabular}{|c|c|c|c|}
\hline
\textbf{Basis Set} & \textbf{HF} & \textbf{GNOF} & \textbf{GNOFm} \\ \hline
\hline
cc-pwCVDZ & 1.755 & 0.037 & 0.039 \\ \hline
cc-pwCVTZ & 1.755 & 0.026 & 0.027 \\ \hline
cc-pwCVQZ & 1.755 & 0.027 & 0.025 \\ \hline
\end{tabular}
\caption{\label{table2} Root mean square deviations (in a.u.) from experimental non-zero molecular dipole moment values computed using Hartree-Fock, GNOF and GNOFm, for Dunning's cc-pwCVDZ, cc-pwCVTZ and cc-pwCVQZ basis sets. Results for each molecule are given in the Supplementary Information, together with bibliographic references corresponding to experimental data.}
\end{table}

Finally, in Table \ref{table2} we report root mean square deviations (in a.u.) from experimental data corresponding to non-zero molecular dipole moments. Results for each molecular system, method, and basis set, can be found in the Supplementary Information, as well as literature sources corresponding to experimental data. HF provides same errors for all studied basis sets. In fact, increasing the basis reduces representation error in HF, but it does not add correlation. Consequently, HF converges smoothly to the HF limit of dipole moments, which can differ from the correlated limit by a nearly basis-independent offset \cite{dip-hartreefock}. Previous studies \cite{Mitxelena2016} showed that dipole, quadrupole and octupole moments computed with electron-pairing-based NOFs compare with CCSD and MRSD-CI for small molecular systems. According to the results shown in Table \ref{table2}, both GNOF and GNOFm improve significantly HF dipole moments, and their errors reduce with increasing the basis set, particularly for GNOFm.

\section{Closing remarks}\label{sec:closing}

This study assesses the performance of the most recent electron-pairing-based natural orbital functionals, GNOF and its modified variant GNOFm, on absolute correlation energies for five- and six-membered rings. This benchmark set, composed of simple aromatic rings of broad relevance, has previously been used to examine the performance and convergence properties of the M{\o}ller–Plesset series and coupled-cluster methods (including iterative approximations). Our results show that GNOFm attains qualitative agreement with the ground-state reference CCSD(T) across multiple sizes of Dunning's weighted core-valence cc-pwCVNZ basis sets. We also report complete-basis-set (CBS) extrapolated correlation energies for GNOF, GNOFm, and CCSD(T). A direct comparison between GNOFm and CCSD(T) indicates agreement to approximately 50 mE$_{h}$, suggesting that GNOFm can be employed to describe systems with dynamic electron correlation effects, in contrast to previous electron-pairing based NOFs.

The error analysis and CBS extrapolations reported here for a representative set of five- and six-membered molecules clarify the capabilities and limitations of Global NOFs and help indicate when their application is most practical. While prior NOFT studies have often targeted systems with significant static correlation, the molecules investigated here are dominated by dynamic correlation. In this regime, GNOFm systematically improves upon GNOF in describing electron correlation, yielding a more balanced account of correlation effects.

The benchmarking dataset used here will be extended in a forthcoming work to guide refinements of the functional form, with particular emphasis on improving the treatment of electron correlation within the NOFT framework. Overall, the findings support the continued development of NOFs as a viable alternative to traditional density functionals and multireference wavefunction methods. Their ability to capture both static and dynamic correlation without active-space selection makes them especially attractive for complex chemical systems, particularly when large molecules are involved. Future efforts should focus on further improving the balanced description of dynamic and non-dynamic correlations within NOFT to enhance accuracy across a broader range of chemical environments. In addition, and following the dipole moments reported in Table \ref{table2}, our laboratory is investigating the impact of incorporating dynamic correlation effects directly into the functional on the charge distribution. The latter implies not only electric multipole moments such as dipoles or quadruples, but also equilibrium geometries.

\section*{Data availability}
The materials that support the findings of this study are openly
available at \url{https://osf.io/64pcn/overview?view_only=72e6a504f07042bbb6d7ccf55aaf29c6}.

\section*{Supplementary Information}
Additional calculations carried out with the Dunning's cc-pVXZ (X=2-5) basis sets, as well as a detailed study of dipole moments.

\section{Acknowledgments}
Financial support comes from the Eusko Jaurlaritza (Basque Government), Ref.: IT1584-22 and from the Grant No. PID 2021-126714NB-I00, funded by MCIN/AEI/10.13039/501100011033. J. F. H. Lew-Yee acknowledges the DIPC and the MCIN program ``Severo Ochoa'' under reference AEI/CEX2018-000867-S for post-doctoral funding (Ref.: 2023/74.) The authors acknowledge the technical and human support provided by both the DIPC Supercomputing Center and IZO-SGI (SGIker) of EHU and European funding (ERDF and ESF).



\renewcommand{\thefigure}{S\arabic{figure}}
\renewcommand{\arraystretch}{1.2}
\setcounter{figure}{0}

\section*{Supplementary Material}

\subsection{Calculations with the Dunning's cc-pVXZ basis set}

In the present section, we show correlation energies obtained for the correlation-consistent Dunning's basis sets (cc-pVXZ, X=2-5). The latter are preferable to be used in calculations with frozen core electrons, however, due to their relevance to the historical development of NOF approximations and their widespread use, we include the corresponding correlation energy analysis below.

\begin{figure}[thb]
\begin{centering}
{\includegraphics[scale=0.27]{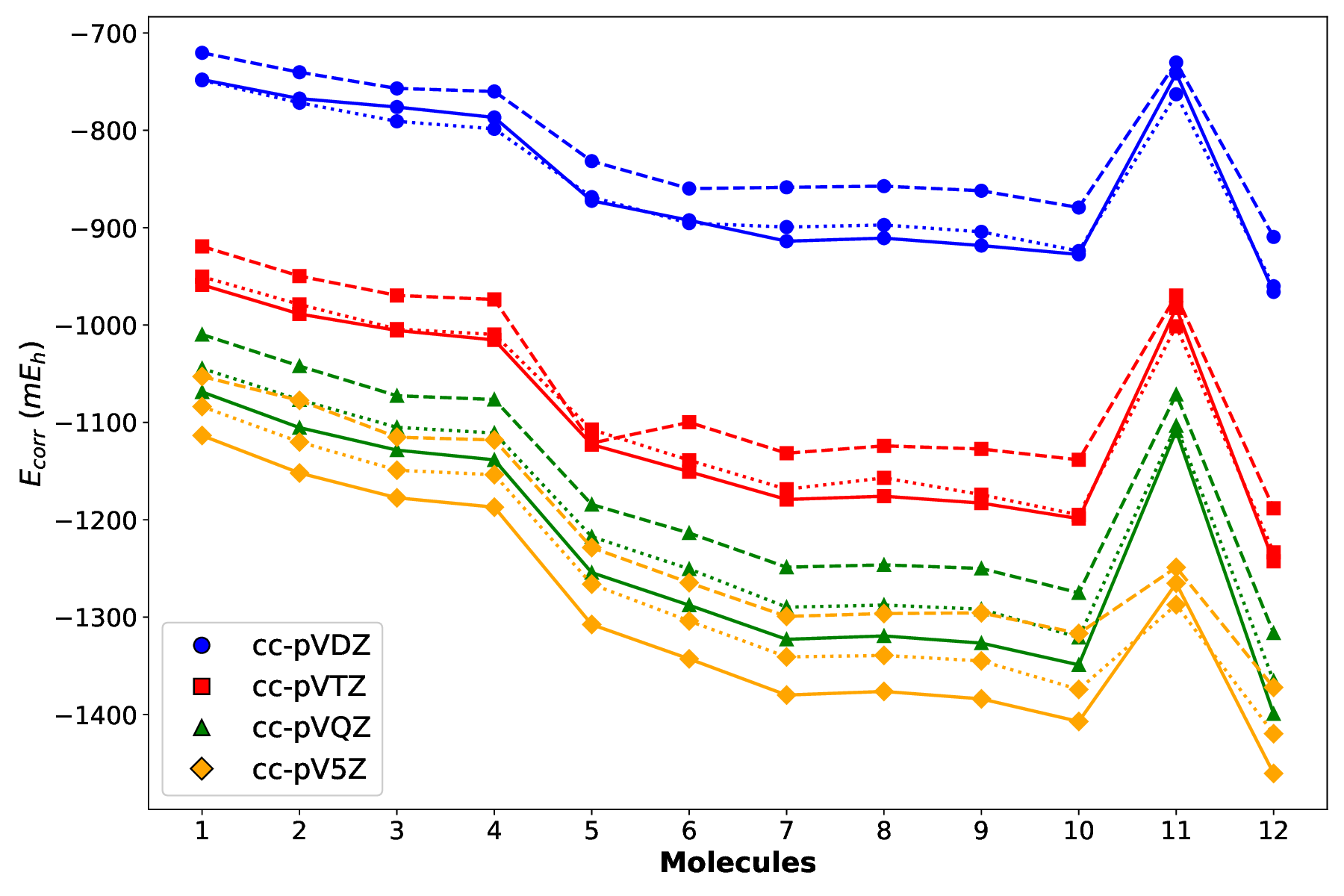}}
\caption{\label{fig:Ecorr_ccd-cc5} Correlation energies ($E-E_\mathrm{HF}$) in mE$_{h}$ for the selected set of molecules, obtained by using GNOF (dashed lines), GNOFm (dotted lines), and CCSD(T) (solid lines) using the cc-pVXZ basis sets. The cardinal numbers X = 2, 3, 4, and 5 correspond to blue, red, green and orange, respectively. Molecules ordered according to the numbering given in Table \ref{table1}.}
\end{centering}
\end{figure}

Correlation energies for GNOF, GNOFm, and CCSD(T) are shown in Fig. \ref{fig:Ecorr_ccd-cc5} for increasing size cc-pVXZ basis sets, being X=2-5 the cardinal number of the basis. Here, molecules are ordered according to the numbering presented in Table \ref{table1}.  NOF and CCSD(T) curves are roughly parallel, so the molecular description agrees for both methods independently of the size of the basis set, as well as of the different studied molecular rings.

\begin{table}[htbp]
\centering
\setlength{\arrayrulewidth}{0.1mm} 
\setlength{\tabcolsep}{2pt} 
\renewcommand{\arraystretch}{1.7} 
\begin{tabular}{|c|c|c|c|c|}
\hline
No. & \textbf{Systems} & \textbf{GNOF} & \textbf{GNOFm} & \textbf{CCSD(T)} \\ \hline
\hline
1 & Cyclopentadiene & -1051.3 & -1084.4 & -1110.6 \\ \hline
2 & Pyrrole & -1081.7 & -1118.9 & -1149.0 \\ \hline
3 & Furan & -1114.7 & -1148.2 & -1173.7 \\ \hline
4 & Imidazole & -1118.3 & -1153.1 & -1183.4 \\ \hline
5 & Benzene & -1245.5 & -1265.8 & -1304.0 \\ \hline
6 & Pyridine & -1262.8 & -1301.7 & -1338.9 \\ \hline
7 & Pyrazine & -1302.9 & -1343.0 & -1375.2 \\ \hline
8 & Pyrimidine & -1298.5 & -1338.3 & -1371.6 \\ \hline
9 & Pyridazine & -1299.9 &-1346.6 & -1379.0 \\ \hline
10 & Triazine & -1320.3 & -1374.8 & -1402.0 \\ \hline
11 & Thiophene & -1179.6 & -1214.3 & -1204.5 \\ \hline
12 & Tetrazine & -1372.4 & -1419.3 & -1453.8 \\ \hline
\end{tabular}
\caption{\label{table1} Complete basis set (CBS) extrapolated correlation energies ($E-E_\mathrm{HF}$) in mE$_{h}$ for the 12 molecular systems, computed using GNOF, GNOFm, and CCSD(T). Helgaker's extrapolation scheme, $E_{\infty} + bX^{-3}$, was employed with $X = 2, 3, 4, 5$ as the cardinal number of the basis set.}
\end{table}

GNOF and GNOFm complete basis set limit (CBS) correlation energy estimates are given in Table \ref{table1}, which also presents CCSD(T) values, performed using the same procedure. Helgaker's extrapolation scheme, $E_{\infty} + bX^{-3}$ with $X = 2, 3, 4, 5$, was employed to carry out the CBS extrapolation. An inspection of CBS limit molecular correlation energies reveals an agreement within 100 mE$_{h}$ between GNOF and CCSD(T), values that are even improved to around 50 mE$_{h}$ when GNOFm is utilized. The results shown in Table \ref{table1} are summarized in Fig. \ref{fig:Ecorr_cbslimit}. GNOFm energies systematically get closer to CCSD(T) results when the static term between electron pairs is modified as described in the section II from the manuscript. In other words, Fig. \ref{fig:Ecorr_cbslimit} reveals that GNOFm CBS correlation energies reduce differences between GNOF and CCSD(T) to the half. Additionally, as it is shown in Fig. \ref{fig:Ecorr_ccd-cc5}, the improvement is obtained for all basis sets studied.

\bigskip

\begin{figure}[thb]
\begin{centering}
{\includegraphics[scale=0.27]{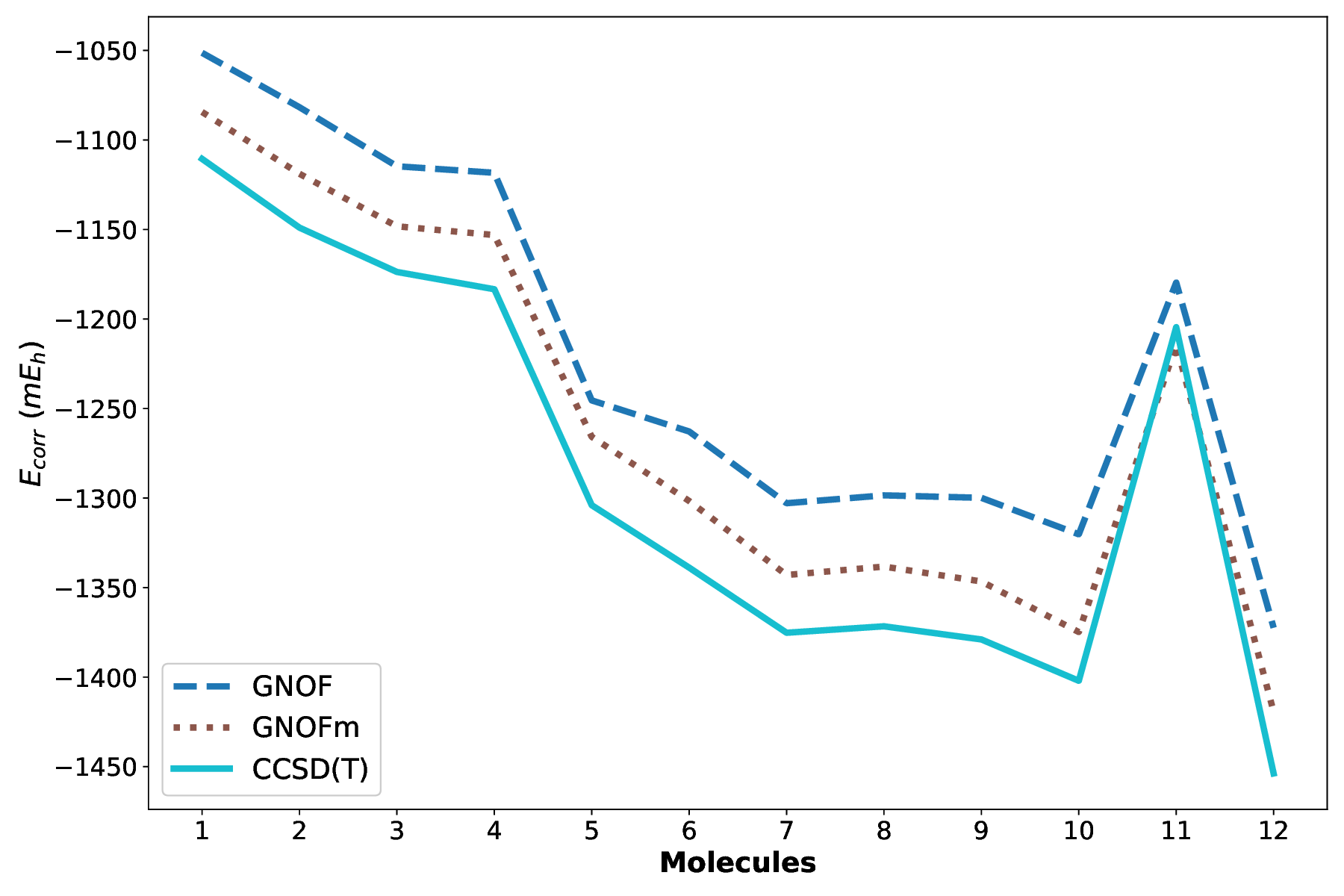}}
\caption{\label{fig:Ecorr_cbslimit} Complete basis set (CBS) extrapolated correlation energies ($E-E_\mathrm{HF}$) in mE$_{h}$ for the 12 molecular systems, computed using GNOF, GNOFm, and CCSD(T). Helgaker's extrapolation scheme, $E_{\infty} + bX^{-3}$, was employed with $X = 2, 3, 4, 5$ as the cardinal number of the basis set.}
\end{centering}
\end{figure}

\subsection{Molecular dipole moments}

In this section, we show the dipole moments obtained by using GNOF and GNOFm approximations, together with reference Hartree-Fock and experimental values. Literature sources corresponding to the latter, as well as numerical values (in a.u.), are given in Table \ref{table2}. Both GNOF and GNOFm improves Hartree-Fock values and corresponding dipole moments get closer to experimental data when the basis set is increased beyond cc-pwCVDZ, as it is shown for Pyrrole, Imidazole, or Pyridazine. GNOFm provides the most accurate values for the largest cc-pwCVQZ basis set, whereas GNOF dipole moments are similar, in average, for the cc-pwCVTZ and cc-pwCVQZ basis sets. Thiophene molecule corresponds to largest errors, and the latter do not improve if $d$ functions are removed from the basis set. In fact, corresponding dipole moments for the cc-pwCVQZ basis set read as 0.2703 a.u. and 0.2833 a.u., respectively for GNOF and GNOFm, which do not change significantly the errors obtained with the aug-cc-pwCVQZ basis set as it is shown in Table \ref{table2}. HF gives same values regardless of the cardinal number of the basis set, as discussed in the manuscript.

\begin{table*}[htbp]
\centering
\setlength{\arrayrulewidth}{0.4mm} 
\setlength{\tabcolsep}{4pt} 
\renewcommand{\arraystretch}{1.7} 
\begin{tabular}{|c|c|c|c|c|c|c|c|c|}
\hline
\textbf{Systems} & \textbf{1} & \textbf{2} & \textbf{3} & \textbf{4} & \textbf{6} & \textbf{8} & \textbf{9} & \textbf{11} \\ \hline
\hline
HF/cc-pwCVXZ & 1.222 & 0.032 & 1.914 & 1.965 & 2.213 & 2.242 & 3.823 & 3.662 \\ \hline
GNOF/cc-pwCVDZ & 0.160 & 0.776 & 0.270 & 1.480 & 0.851 & 0.904 & 1.639 & 0.258 \\ \hline
GNOF/cc-pwCVTZ & 0.159 & 0.737 & 0.283 & 1.475 & 0.870 & 0.910 & 1.655 & 0.266 \\ \hline
GNOF/cc-pwCVQZ & 0.165 & 0.727 & 0.291 & 1.476 & 0.885 & 0.913 & 1.663 & 0.271 \\ \hline
GNOFm/cc-pwCVDZ & 0.163 & 0.762 & 0.271 & 1.470 & 0.847 & 0.893 & 1.627 & 0.283 \\ \hline
GNOFm/cc-pwCVTZ & 0.159 & 0.731 & 0.282 & 1.463 & 0.856 & 0.895 & 1.646 & 0.270 \\ \hline
GNOFm/cc-pwCVQZ & 0.167 & 0.724 & 0.286 & 1.462 & 0.886 & 0.904 & 1.656 & 0.271 \\ \hline
\textbf{Exp.} & 0.165 & 0.695 & 0.268 & 1.444 & 0.871 & 0.918 & 1.660 & 0.216 \\ \hline
\textbf{Ref.} & (\citenum{dip-cyclo}) & (\citenum{dip-pyrrole}) & (\citenum{dip-furan}) & (\citenum{dip-imidazole}) & (\citenum{dip-pyridine}) & (\citenum{dip-pyrimidine}) & (\citenum{dip-pyridazine}) & (\citenum{dip-thiophene}) \\ \hline
\end{tabular}
\caption{\label{table2} Non-zero dipole moments (in a.u.) computed using HF, GNOF and GNOFm with Dunning's cc-pwCVXZ basis sets ($X = 2, 3, 4$), together with experimental values and their corresponding reference from the literature. Results corresponding to thiophene (no. 11) are obtained with the aug-cc-pwCVXZ basis sets. Molecular numbering reads as follows: 1. Cyclopentadiene, 2. Pyrrole, 3. Furan, 4. Imidazole, 6. Pyridine, 8. Pyrimidine, 9. Pyridazine, 11. Thiophene.}
\end{table*}

\providecommand{\latin}[1]{#1}
\makeatletter
\providecommand{\doi}
  {\begingroup\let\do\@makeother\dospecials
  \catcode`\{=1 \catcode`\}=2 \doi@aux}
\providecommand{\doi@aux}[1]{\endgroup\texttt{#1}}
\makeatother
\providecommand*\mcitethebibliography{\thebibliography}
\csname @ifundefined\endcsname{endmcitethebibliography}  {\let\endmcitethebibliography\endthebibliography}{}

\end{document}